\documentclass[11pt, letterpaper, onecolumn, peerreview]{IEEEtran}
\usepackage{exscale}
\usepackage{color}
\usepackage{colordvi}
\usepackage{graphicx}
\usepackage{amsmath}
\usepackage{amssymb}
\usepackage{epic}

\newcommand{\real}{I\!\!R}

\newtheorem{theorem}{Theorem}[section]

\newtheorem{definition}[theorem]{Definition}

\title{\large \bf Stabilization using both noisy and noiseless
  feedback} 

\author{Anant Sahai\footnote{Prof.~A.~Sahai is with the Wireless
Foundations Center within the Department of Electrical Engineering and
Computer Science at the University of California at Berkeley. This
work was initially presented at MTNS in Kyoto in 2006.}  \\ {\small
sahai@eecs.berkeley.edu}}

\begin{document}

\maketitle

\begin{abstract} 
  When designing a distributed control system, the system designer has
  a choice in how to connect the different units through communication
  channels. In practice, noiseless and noisy channels may coexist.
  Using the standard toy example of scalar stabilization, this paper
  shows how a small amount of noiseless feedback can perform a
  ``supervisory'' role and thereby boost the effectiveness of noisy
  feedback.
\end{abstract}


\IEEEpeerreviewmaketitle

\section{Introduction}

Distributed control has long been understood as a challenging problem,
and in such situations the control signals often have a dual purpose
--- to achieve the control objectives as well as to communicate
necessary information between the distributed systems
\cite{Witsenhausen68, WitsenhausenPatterns,
AreaExamPaper}. Consequently, it had been recognized that there may
lie deep connections between information theory and control theory
\cite{HoKastnerWong}, but establishing such connections remained out
of reach. Recently however, some deep connections have begun to
emerge, most notably in the context of stabilization over a limited
communication medium. The reader is directed to the recent September
2004 issue of IEEE Transactions on Automatic Control and the articles
(and references) therein for a more comprehensive survey, but it is
worthwhile to make note of some relevant prior work here.

The first truly information-theoretic connection was given by
Tatikonda with his work on sequential rate-distortion theory. This
married the causality constraint critical in control problems to the
rate-distortion framework provided by Shannon. By using Massey's
directed mutual information and relaxing the optimization to be over
all channels with the same directed mutual information instead of the
original channel, it provides a lower-bound on the achievable
closed-loop performance of a control system over a channel, noisy or
otherwise. Because this bound is sometimes infinite, it also implies
that there is a fundamental rate of information production, namely the
sum of the logs of the unstable eigenvalues of the plant, that is
attached to an unstable linear discrete-time process
\cite{TatikondaThesis, Tatikonda1}. The corresponding achievability
result was based on noiseless, but rate-limited, channels. Nair et
al.~extended the noiseless rate-limited story to the case of unbounded
disturbances and observation noise under suitable conditions
\cite{NairPaper1, NairPaper2}. 

The noisy channel side of the story is distinct and involves
additional subtleties. Much work was tied to specific communication
channels. We had previously showed that it is possible to stabilize
persistently disturbed controlled Gauss-Markov processes over suitable
power-constrained AWGN (Additive White Gaussian Noise)
channels\cite{OurACC99Paper, OurMainLQGPaper} where it turns out that
Shannon capacity is tight and linear observers and controllers are
sufficient to achieve stabilization \cite{Bansal}. In contrast, we
showed that the Shannon capacity of the binary erasure channel (BEC)
is not sufficient to check stabilizability and introduced the anytime
capacity as a candidate figure of merit in
\cite{ACC00Paper}. Following up on our treatment of the BEC case,
Martins et al.~have studied more general erasure-type models and have
also incorporated bounded model uncertainty in the plant
\cite{Nuno}. Elia used ideas from robust control to deal with
communication uncertainty in a mixed continuous/discrete context, but
restricting to linear operations \cite{EliaPaper1, EliaPaper2}.

Our work in \cite{ControlPartI, SahaiThesis} established another deep
connection in the noisy channel case by showing how stabilization
problems over noisy channels are equivalent to anytime communication
problems where the encoder has access to feedback. These results are
extended in \cite{ControlPartII} to the vector case. In stabilization
problems, the sense of stability desired (mean-squared, fourth-moment,
etc.) determines the target anytime reliability in conjunction with
the unstable eigenvalues. In contrast, the rate requirement comes only
from the unstable eigenvalues just as it does in the noiseless
rate-limited channel case.

The feedback anytime capacity is lower bounded by the anytime capacity
without feedback in \cite{OurSourceCodingPaper, SahaiThesis}. The
basic result is that the block random-coding error exponent $E_r(R)$
is a lower bound to the anytime reliability at rate $R$. Furthermore,
implementable observers and controllers can be designed that approach
this lower bound by using a nearly memoryless quantization strategy
for the observer and a stack-algorithm based strategy for the
controller \cite{Allerton05Control, HariThesis}. However, using purely
information-theoretic arguments \cite{OurUpperBoundPaper} shows that
the anytime reliability with feedback can be much higher and provides
a new upper bound called the ``focusing bound.'' For symmetric
channels, the focusing bound says that at rate $R =
\frac{E_0(\rho)}{\rho}$, the anytime reliability can not exceed
$E_0(\rho)$ where $E_0(\rho)$ is the Gallager function defined as:
\begin{equation} \label{eqn:enought}
  E_0(\rho) = \max_{\vec{q}} -\ln \sum_y \bigg[ \sum_x q_x
  p_{x,y}^\frac{1}{1+\rho} \bigg]^{(1+\rho)}
\end{equation}
Note that for symmetric channels, it suffices to use a uniform
$\vec{q}$ while optimizing (\ref{eqn:enought}) \cite{Gallager}. 

\cite{OurUpperBoundPaper} also shows that the focusing bound is
asymptotically achievable for any erasure type channel or noisy
channel that has a strictly positive zero-error capacity. (For
non-erasure channels whose zero-error capacity is zero, the
achievability of the focusing bound is still open, although it is
known that we can beat the lower-bound of $E_r(R)$ in essentially all
cases \cite{MyITA06Paper, TuncThesis, SimsekJainVaraiya}.) The achievability
strategy of \cite{OurUpperBoundPaper} is interesting because it has
strictly bounded computation at both the encoder and decoder and
furthermore, it relies on sending very low-rate ``supervisory''
messages using the ``noiseless'' aspects of the communication channel.

While it is possible to directly apply the results of
\cite{OurUpperBoundPaper} to the stabilization problem by means of the
equivalence established in \cite{ControlPartI} and a separation
architecture, it is more illuminating to give the strategy directly in
the stabilization context without reference to anytime communication
per-se. After setting up some notation and preliminaries in
Section~\ref{sec:preliminaries}, we give the stabilization scheme in
Section~\ref{sec:mainresult}, before concluding in
Section~\ref{sec:conclusion}. Proof ideas are given here,
but the ``heavy lifting'' is found in \cite{ControlPartI,
OurUpperBoundPaper}.

\section{Preliminaries} \label{sec:preliminaries}
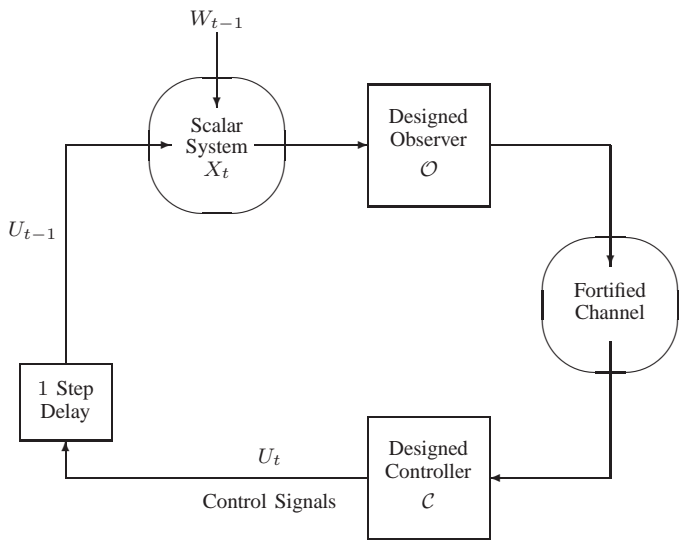
\begin{figure}
\begin{center}
\setlength{\unitlength}{2500sp}%
\begingroup\makeatletter\ifx\SetFigFont\undefined%
\gdef\SetFigFont#1#2#3#4#5{%
  \reset@font\fontsize{#1}{#2pt}%
  \fontfamily{#3}\fontseries{#4}\fontshape{#5}%
  \selectfont}%
\fi\endgroup%
\begin{picture}(7643,5307)(95,-4498)
\thinlines
\put(301,-3511){\framebox(900,750){}}
\put(751,-3061){\makebox(0,0)[b]{\smash{\SetFigFont{8}{14.4}{\rmdefault}{\mddefault}{\updefault}$1$ Step}}}
\put(751,-3286){\makebox(0,0)[b]{\smash{\SetFigFont{8}{14.4}{\rmdefault}{\mddefault}{\updefault}Delay}}}
\put(6150,-2169){\oval(1342,1342)}
\put(6151,-2311){\makebox(0,0)[b]{\smash{\SetFigFont{8}{14.4}{\rmdefault}{\mddefault}{\updefault}Channel}}}
\put(6151,-2086){\makebox(0,0)[b]{\smash{\SetFigFont{8}{14.4}{\rmdefault}{\mddefault}{\updefault}Fortified}}}
\put(4351,-361){\makebox(0,0)[b]{\smash{\SetFigFont{8}{14.4}{\rmdefault}{\mddefault}{\updefault}Designed}}}
\put(4351,-586){\makebox(0,0)[b]{\smash{\SetFigFont{8}{14.4}{\rmdefault}{\mddefault}{\updefault}Observer}}}
\put(4351,-3661){\makebox(0,0)[b]{\smash{\SetFigFont{8}{14.4}{\rmdefault}{\mddefault}{\updefault}Designed}}}
\put(4351,-3886){\makebox(0,0)[b]{\smash{\SetFigFont{8}{14.4}{\rmdefault}{\mddefault}{\updefault}Controller}}}
\put(2251,-586){\oval(1342,1342)}
\put(3751,-1186){\framebox(1200,1200){}}
\put(3751,-4486){\framebox(1200,1200){}}
\put(2626,-586){\vector( 1, 0){1125}}
\put(2251,539){\line( 0,-1){525}}
\put(2251, 14){\vector( 0,-1){225}}
\put(4951,-586){\line( 1, 0){1200}}
\put(6151,-586){\vector( 0,-1){1200}}
\put(3751,-3886){\line(-1, 0){3000}}
\put(751,-3886){\vector( 0, 1){375}}
\put(751,-2761){\line( 0, 1){2175}}
\put(751,-586){\vector( 1, 0){1050}}
\put(2776,-3736){\makebox(0,0)[b]{\smash{\SetFigFont{8}{14.4}{\rmdefault}{\mddefault}{\updefault}$U_t$}}}
\put(2776,-4186){\makebox(0,0)[b]{\smash{\SetFigFont{8}{14.4}{\rmdefault}{\mddefault}{\updefault}Control Signals}}}
\put(4351,-886){\makebox(0,0)[b]{\smash{\SetFigFont{8}{14.4}{\rmdefault}{\mddefault}{\updefault}$\cal{O}$}}}
\put(4351,-4186){\makebox(0,0)[b]{\smash{\SetFigFont{8}{14.4}{\rmdefault}{\mddefault}{\updefault}$\cal{C}$}}}
\put(2251,-886){\makebox(0,0)[b]{\smash{\SetFigFont{8}{14.4}{\rmdefault}{\mddefault}{\updefault}$X_t$}}}
\put(2251,614){\makebox(0,0)[b]{\smash{\SetFigFont{8}{14.4}{\rmdefault}{\mddefault}{\updefault}$W_{t-1}$}}}
\put(426,-1486){\makebox(0,0)[b]{\smash{\SetFigFont{8}{14.4}{\rmdefault}{\mddefault}{\updefault}$U_{t-1}$}}}
\put(2251,-451){\makebox(0,0)[b]{\smash{\SetFigFont{8}{14.4}{\rmdefault}{\mddefault}{\updefault}Scalar}}}
\put(2251,-661){\makebox(0,0)[b]{\smash{\SetFigFont{8}{14.4}{\rmdefault}{\mddefault}{\updefault}System}}}
\put(6151,-3886){\vector(-1, 0){1200}}
\put(6151,-2536){\line( 0,-1){1350}}
\end{picture}
\end{center}
\caption{Control over a fortified communication channel.}
\label{fig:nofeedbackatallproblem}
\end{figure}

The scalar stabilization problem is illustrated in
Figure~\ref{fig:nofeedbackatallproblem}. 

\begin{equation} \label{eqn:discretesystem} 
X_{t+1} = \lambda X_{t} + U_{t} + W_{t}, \ \ t \geq 0
\end{equation} 
where $\{ X_{t} \}$ is a ${\real}$-valued state process. $\{ U_{t} \}$
is a ${\real}$-valued control process and $\{ W_{t} \}$ is a bounded
noise/disturbance process s.t.~$|W_t| \leq \frac{\Omega}{2}$. This
bound is assumed to hold with certainty. For convenience, we also
assume a known initial condition $X_0=0$. To make things interesting,
consider $\lambda > 1$ so the open-loop system is exponentially
unstable. The observer/encoder system $\cal O$ observes $X_t$ and
generates inputs $a_t$ to the channel. The decoder/controller system
$\cal C$ observes channel outputs $B_t$ and generates control signals
$U_t$. Both $\cal O, C$ are allowed to have unbounded memory and to be
nonlinear in general. The goal is stability:

\begin{definition} \label{def:momentstable}
A closed-loop dynamic system with state $X_t$ is {\em $\eta$-stable}
if there exists a constant $K$ s.t.~$E[|X_t|^\eta] \leq K$ for all
$t \geq 0$.
\end{definition}
\vspace{0.1in}

\begin{definition} 
A {\em discrete time discrete memoryless channel} (DMC) is a
probabilistic system with an input. At every time step $t$, it takes
an input $x_t \in {\cal X}$ and produces an output $y_t \in {\cal Y}$
with probability $p(y_t|x_t)$. Both ${\cal X},{\cal Y}$ are finite
sets. The current channel output is independent of all past random
variables in the system conditioned on the current channel input.
\end{definition}
\vspace{0.1in}

The twist in this paper is that the noisy channel does not have be
used alone. In addition, we have a low-rate noiseless channel to help
out. 

\begin{definition} \label{def:fortify}
Given a DMC $P$ for the forward link, a $\frac{1}{k}$-{\em fortified}
channel built around it is one in which every $k$-th use of $P$ is
supplemented with the ability to transmit a single noise-free bit to
the receiver.
\end{definition}
\vspace{0.1in} 

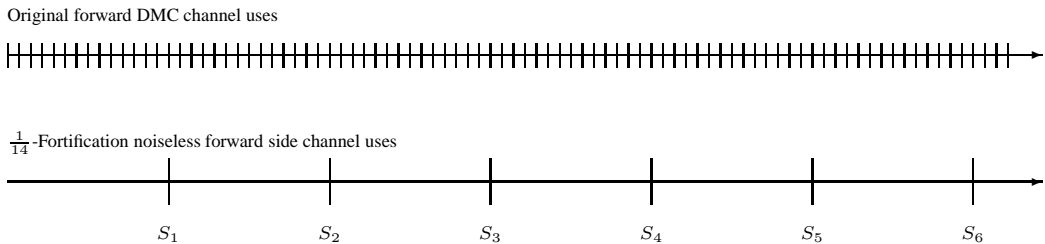
\begin{figure}
\begin{center}
\setlength{\unitlength}{3800sp}%
\begingroup\makeatletter\ifx\SetFigFont\undefined%
\gdef\SetFigFont#1#2#3#4#5{%
  \reset@font\fontsize{#1}{#2pt}%
  \fontfamily{#3}\fontseries{#4}\fontshape{#5}%
  \selectfont}%
\fi\endgroup%
\begin{picture}(6774,1627)(-11,-1469)
\thinlines
{\color[rgb]{0,0,0}\put(  1,-136){\line( 0,-1){150}}
}%
{\color[rgb]{0,0,0}\put( 76,-136){\line( 0,-1){150}}
}%
{\color[rgb]{0,0,0}\put(151,-136){\line( 0,-1){150}}
}%
{\color[rgb]{0,0,0}\put(226,-136){\line( 0,-1){150}}
}%
{\color[rgb]{0,0,0}\put(301,-136){\line( 0,-1){150}}
}%
{\color[rgb]{0,0,0}\put(376,-136){\line( 0,-1){150}}
}%
{\color[rgb]{0,0,0}\put(451,-136){\line( 0,-1){150}}
}%
{\color[rgb]{0,0,0}\put(526,-136){\line( 0,-1){150}}
}%
{\color[rgb]{0,0,0}\put(601,-136){\line( 0,-1){150}}
}%
{\color[rgb]{0,0,0}\put(676,-136){\line( 0,-1){150}}
}%
{\color[rgb]{0,0,0}\put(751,-136){\line( 0,-1){150}}
}%
{\color[rgb]{0,0,0}\put(826,-136){\line( 0,-1){150}}
}%
{\color[rgb]{0,0,0}\put(901,-136){\line( 0,-1){150}}
}%
{\color[rgb]{0,0,0}\put(976,-136){\line( 0,-1){150}}
}%
{\color[rgb]{0,0,0}\put(1051,-136){\line( 0,-1){150}}
}%
{\color[rgb]{0,0,0}\put(1126,-136){\line( 0,-1){150}}
}%
{\color[rgb]{0,0,0}\put(1201,-136){\line( 0,-1){150}}
}%
{\color[rgb]{0,0,0}\put(1276,-136){\line( 0,-1){150}}
}%
{\color[rgb]{0,0,0}\put(1351,-136){\line( 0,-1){150}}
}%
{\color[rgb]{0,0,0}\put(1426,-136){\line( 0,-1){150}}
}%
{\color[rgb]{0,0,0}\put(1501,-136){\line( 0,-1){150}}
}%
{\color[rgb]{0,0,0}\put(1576,-136){\line( 0,-1){150}}
}%
{\color[rgb]{0,0,0}\put(1651,-136){\line( 0,-1){150}}
}%
{\color[rgb]{0,0,0}\put(1726,-136){\line( 0,-1){150}}
}%
{\color[rgb]{0,0,0}\put(1801,-136){\line( 0,-1){150}}
}%
{\color[rgb]{0,0,0}\put(1876,-136){\line( 0,-1){150}}
}%
{\color[rgb]{0,0,0}\put(1951,-136){\line( 0,-1){150}}
}%
{\color[rgb]{0,0,0}\put(2026,-136){\line( 0,-1){150}}
}%
{\color[rgb]{0,0,0}\put(2101,-136){\line( 0,-1){150}}
}%
{\color[rgb]{0,0,0}\put(2176,-136){\line( 0,-1){150}}
}%
{\color[rgb]{0,0,0}\put(2251,-136){\line( 0,-1){150}}
}%
{\color[rgb]{0,0,0}\put(2326,-136){\line( 0,-1){150}}
}%
{\color[rgb]{0,0,0}\put(2401,-136){\line( 0,-1){150}}
}%
{\color[rgb]{0,0,0}\put(2476,-136){\line( 0,-1){150}}
}%
{\color[rgb]{0,0,0}\put(2551,-136){\line( 0,-1){150}}
}%
{\color[rgb]{0,0,0}\put(2626,-136){\line( 0,-1){150}}
}%
{\color[rgb]{0,0,0}\put(2701,-136){\line( 0,-1){150}}
}%
{\color[rgb]{0,0,0}\put(2776,-136){\line( 0,-1){150}}
}%
{\color[rgb]{0,0,0}\put(2851,-136){\line( 0,-1){150}}
}%
{\color[rgb]{0,0,0}\put(2926,-136){\line( 0,-1){150}}
}%
{\color[rgb]{0,0,0}\put(3001,-136){\line( 0,-1){150}}
}%
{\color[rgb]{0,0,0}\put(3076,-136){\line( 0,-1){150}}
}%
{\color[rgb]{0,0,0}\put(3151,-136){\line( 0,-1){150}}
}%
{\color[rgb]{0,0,0}\put(3226,-136){\line( 0,-1){150}}
}%
{\color[rgb]{0,0,0}\put(3301,-136){\line( 0,-1){150}}
}%
{\color[rgb]{0,0,0}\put(3376,-136){\line( 0,-1){150}}
}%
{\color[rgb]{0,0,0}\put(3451,-136){\line( 0,-1){150}}
}%
{\color[rgb]{0,0,0}\put(3526,-136){\line( 0,-1){150}}
}%
{\color[rgb]{0,0,0}\put(3601,-136){\line( 0,-1){150}}
}%
{\color[rgb]{0,0,0}\put(3676,-136){\line( 0,-1){150}}
}%
{\color[rgb]{0,0,0}\put(3751,-136){\line( 0,-1){150}}
}%
{\color[rgb]{0,0,0}\put(3826,-136){\line( 0,-1){150}}
}%
{\color[rgb]{0,0,0}\put(3901,-136){\line( 0,-1){150}}
}%
{\color[rgb]{0,0,0}\put(3976,-136){\line( 0,-1){150}}
}%
{\color[rgb]{0,0,0}\put(4051,-136){\line( 0,-1){150}}
}%
{\color[rgb]{0,0,0}\put(4126,-136){\line( 0,-1){150}}
}%
{\color[rgb]{0,0,0}\put(4201,-136){\line( 0,-1){150}}
}%
{\color[rgb]{0,0,0}\put(4276,-136){\line( 0,-1){150}}
}%
{\color[rgb]{0,0,0}\put(4351,-136){\line( 0,-1){150}}
}%
{\color[rgb]{0,0,0}\put(4426,-136){\line( 0,-1){150}}
}%
{\color[rgb]{0,0,0}\put(4501,-136){\line( 0,-1){150}}
}%
{\color[rgb]{0,0,0}\put(4576,-136){\line( 0,-1){150}}
}%
{\color[rgb]{0,0,0}\put(4651,-136){\line( 0,-1){150}}
}%
{\color[rgb]{0,0,0}\put(4726,-136){\line( 0,-1){150}}
}%
{\color[rgb]{0,0,0}\put(4801,-136){\line( 0,-1){150}}
}%
{\color[rgb]{0,0,0}\put(4876,-136){\line( 0,-1){150}}
}%
{\color[rgb]{0,0,0}\put(4951,-136){\line( 0,-1){150}}
}%
{\color[rgb]{0,0,0}\put(5026,-136){\line( 0,-1){150}}
}%
{\color[rgb]{0,0,0}\put(5101,-136){\line( 0,-1){150}}
}%
{\color[rgb]{0,0,0}\put(5176,-136){\line( 0,-1){150}}
}%
{\color[rgb]{0,0,0}\put(5251,-136){\line( 0,-1){150}}
}%
{\color[rgb]{0,0,0}\put(5326,-136){\line( 0,-1){150}}
}%
{\color[rgb]{0,0,0}\put(5401,-136){\line( 0,-1){150}}
}%
{\color[rgb]{0,0,0}\put(5476,-136){\line( 0,-1){150}}
}%
{\color[rgb]{0,0,0}\put(5551,-136){\line( 0,-1){150}}
}%
{\color[rgb]{0,0,0}\put(5626,-136){\line( 0,-1){150}}
}%
{\color[rgb]{0,0,0}\put(5701,-136){\line( 0,-1){150}}
}%
{\color[rgb]{0,0,0}\put(5776,-136){\line( 0,-1){150}}
}%
{\color[rgb]{0,0,0}\put(5851,-136){\line( 0,-1){150}}
}%
{\color[rgb]{0,0,0}\put(5926,-136){\line( 0,-1){150}}
}%
{\color[rgb]{0,0,0}\put(6001,-136){\line( 0,-1){150}}
}%
{\color[rgb]{0,0,0}\put(6076,-136){\line( 0,-1){150}}
}%
{\color[rgb]{0,0,0}\put(6151,-136){\line( 0,-1){150}}
}%
{\color[rgb]{0,0,0}\put(6226,-136){\line( 0,-1){150}}
}%
{\color[rgb]{0,0,0}\put(6301,-136){\line( 0,-1){150}}
}%
{\color[rgb]{0,0,0}\put(6376,-136){\line( 0,-1){150}}
}%
{\color[rgb]{0,0,0}\put(6451,-136){\line( 0,-1){150}}
}%
{\color[rgb]{0,0,0}\put(6526,-136){\line( 0,-1){150}}
}%
{\color[rgb]{0,0,0}\put(  1,-1036){\vector( 1, 0){6750}}
}%
{\color[rgb]{0,0,0}\put(1051,-886){\line( 0,-1){300}}
}%
{\color[rgb]{0,0,0}\put(2101,-886){\line( 0,-1){300}}
}%
{\color[rgb]{0,0,0}\put(3151,-886){\line( 0,-1){300}}
}%
{\color[rgb]{0,0,0}\put(4201,-886){\line( 0,-1){300}}
}%
{\color[rgb]{0,0,0}\put(5251,-886){\line( 0,-1){300}}
}%
{\color[rgb]{0,0,0}\put(6301,-886){\line( 0,-1){300}}
}%
{\color[rgb]{0,0,0}\put(  1,-211){\vector( 1, 0){6750}}
}%
\put(  1, 14){\makebox(0,0)[lb]{\smash{\SetFigFont{7}{6}{\rmdefault}{\mddefault}{\updefault}{\color[rgb]{0,0,0}Original forward DMC channel uses}%
}}}
\put(1051,-1411){\makebox(0,0)[b]{\smash{\SetFigFont{7}{6}{\rmdefault}{\mddefault}{\updefault}{\color[rgb]{0,0,0}$S_1$}%
}}}
\put(2101,-1411){\makebox(0,0)[b]{\smash{\SetFigFont{7}{6}{\rmdefault}{\mddefault}{\updefault}{\color[rgb]{0,0,0}$S_2$}%
}}}
\put(3151,-1411){\makebox(0,0)[b]{\smash{\SetFigFont{7}{6}{\rmdefault}{\mddefault}{\updefault}{\color[rgb]{0,0,0}$S_3$}%
}}}
\put(4201,-1411){\makebox(0,0)[b]{\smash{\SetFigFont{7}{6}{\rmdefault}{\mddefault}{\updefault}{\color[rgb]{0,0,0}$S_4$}%
}}}
\put(5251,-1411){\makebox(0,0)[b]{\smash{\SetFigFont{7}{6}{\rmdefault}{\mddefault}{\updefault}{\color[rgb]{0,0,0}$S_5$}%
}}}
\put(6301,-1411){\makebox(0,0)[b]{\smash{\SetFigFont{7}{6}{\rmdefault}{\mddefault}{\updefault}{\color[rgb]{0,0,0}$S_6$}%
}}}
\put(  1,-811){\makebox(0,0)[lb]{\smash{\SetFigFont{7}{6}{\rmdefault}{\mddefault}{\updefault}{\color[rgb]{0,0,0}$\frac{1}{14}$-Fortification noiseless forward side channel uses}%
}}}
\end{picture}
\end{center}
\caption{Fortification illustrated: the forward noisy channel uses are
supplemented with regular low-rate use of a noiseless side channel.}
\label{fig:fortification}
\end{figure}

\section{Main Result} \label{sec:mainresult}

The main result of this paper is that the focusing bound induced
fundamental limits are asymptotically achievable with even a tiny
amount of fortification. 

\begin{theorem} \label{thm:main}
For a symmetric noisy memoryless channel and any amount of fortification $0 < k < \infty$, it is possible to 
control the unstable scalar plant of (\ref{eqn:discretesystem}) over
that fortified channel so that the $\eta$-moment of $|X_t|$ stays
finite for all time if there exists any $\epsilon > 0$ so that
$\frac{E_0(\eta + \epsilon)}{\eta + \epsilon} > \ln \lambda$ even if
the observer is only allowed to observe the state $X_t$ corrupted by
bounded noise.
\end{theorem}
{\em Proof:} The statement of the theorem does not refer to the
focusing bound, but the connection can be seen by defining $\rho' =
\eta + \epsilon$. Then $\frac{E_0(\rho')}{\rho'} > \ln \lambda$ is
just a statement about the implicit rate $R' =
\frac{E_0(\rho')}{\rho'}$ being high enough. By multiplying both sides
by $\rho'$, it is saying that the reliability $E_0(\rho') > \rho' \ln
\lambda > \eta \ln \lambda$ which is what is necessary for
$\eta$-stabilization. Thus, this theorem is essentially saying that
the focusing bound can be asymptotically achieved with fortification.

The control strategy itself is based on three ideas:
\begin{itemize}
 \item Feedback of channel outputs to the observer by 
       making the plant ``dance'' in an observable way following
       \cite{ControlPartI}.  This makes the proof simpler to
       understand, but is later eliminated by better leveraging the
       next idea:

 \item Nonuniform sampling of the plant state at the observer for
       purposes of ``codeword'' generation for the noisy channel
       uses. The codewords themselves are generated using random
       coding. The plant state is resampled after the controller
       applies a ``true'' (as opposed to ``dancing'') control.

 \item Supervisory use of the noiseless fortification channel by the
       observer to tell the controller how and when to apply a
       ``true'' control.
\end{itemize}

They are covered in greater detail in the next few subsections.

\subsection{Communication through ``dancing''} \label{sec:dancing}

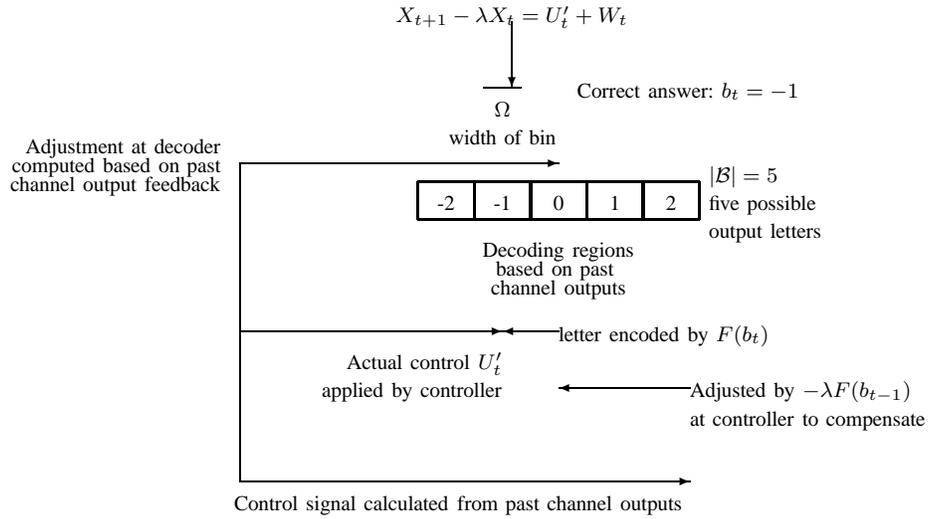
\begin{figure}
\begin{center}
\setlength{\unitlength}{3100sp}%
\begingroup\makeatletter\ifx\SetFigFont\undefined%
\gdef\SetFigFont#1#2#3#4#5{%
  \reset@font\fontsize{#1}{#2pt}%
  \fontfamily{#3}\fontseries{#4}\fontshape{#5}%
  \selectfont}%
\fi\endgroup%
\begin{picture}(3900,4110)(2101,-3640)
\thinlines
{\color[rgb]{0,0,0}\put(5851,-2611){\vector(-1, 0){1050}}
}%
{\color[rgb]{0,0,0}\put(4801,-2161){\vector(-1, 0){450}}
}%
{\color[rgb]{0,0,0}\put(4201,-211){\line( 1, 0){300}}
}%
{\color[rgb]{0,0,0}\put(4426,314){\vector( 0,-1){525}}
}%
{\color[rgb]{0,0,0}\put(2251,-811){\line( 0,-1){2550}}
}%
{\color[rgb]{0,0,0}\put(2251,-2161){\vector( 1, 0){2100}}
}%
{\color[rgb]{0,0,0}\put(2251,-3361){\vector( 1, 0){3600}}
}%
{\color[rgb]{0,0,0}\put(2251,-811){\vector( 1, 0){2550}}
}%
\thicklines
{\color[rgb]{0,0,0}\put(3676,-961){\line( 1, 0){2250}}
}%
{\color[rgb]{0,0,0}\put(3676,-1261){\line( 1, 0){2250}}
}%
{\color[rgb]{0,0,0}\put(4126,-961){\line( 0,-1){300}}
}%
{\color[rgb]{0,0,0}\put(4576,-961){\line( 0,-1){300}}
}%
{\color[rgb]{0,0,0}\put(5026,-961){\line( 0,-1){300}}
}%
{\color[rgb]{0,0,0}\put(3676,-961){\line( 0,-1){300}}
}%
{\color[rgb]{0,0,0}\put(5476,-961){\line( 0,-1){300}}
}%
{\color[rgb]{0,0,0}\put(5926,-961){\line( 0,-1){300}}
}%
\put(4801,-1711){\makebox(0,0)[b]{\smash{\SetFigFont{8}{7.2}{\rmdefault}{\mddefault}{\updefault}{\color[rgb]{0,0,0}based on past }%
}}}
\put(4801,-1861){\makebox(0,0)[b]{\smash{\SetFigFont{8}{7.2}{\rmdefault}{\mddefault}{\updefault}{\color[rgb]{0,0,0}channel outputs}%
}}}
\put(4351,-436){\makebox(0,0)[b]{\smash{\SetFigFont{8}{7.2}{\rmdefault}{\mddefault}{\updefault}{\color[rgb]{0,0,0}$\Omega$}%
}}}
\put(4351,-661){\makebox(0,0)[b]{\smash{\SetFigFont{8}{7.2}{\rmdefault}{\mddefault}{\updefault}{\color[rgb]{0,0,0}width of bin}%
}}}
\put(4801,-2236){\makebox(0,0)[lb]{\smash{\SetFigFont{8}{7.2}{\rmdefault}{\mddefault}{\updefault}{\color[rgb]{0,0,0}letter encoded by $F(b_t)$}%
}}}
\put(4351,-2686){\makebox(0,0)[rb]{\smash{\SetFigFont{8}{7.2}{\rmdefault}{\mddefault}{\updefault}{\color[rgb]{0,0,0}applied by controller}%
}}}
\put(5851,-3586){\makebox(0,0)[rb]{\smash{\SetFigFont{8}{7.2}{\rmdefault}{\mddefault}{\updefault}{\color[rgb]{0,0,0}Control signal calculated from past channel outputs }%
}}}
\put(2101,-736){\makebox(0,0)[rb]{\smash{\SetFigFont{8}{7.2}{\rmdefault}{\mddefault}{\updefault}{\color[rgb]{0,0,0}Adjustment at decoder}%
}}}
\put(2101,-886){\makebox(0,0)[rb]{\smash{\SetFigFont{8}{7.2}{\rmdefault}{\mddefault}{\updefault}{\color[rgb]{0,0,0}computed based on past}%
}}}
\put(2101,-1036){\makebox(0,0)[rb]{\smash{\SetFigFont{8}{7.2}{\rmdefault}{\mddefault}{\updefault}{\color[rgb]{0,0,0}channel output feedback}%
}}}
\put(4801,-1186){\makebox(0,0)[b]{\smash{\SetFigFont{8}{7.2}{\rmdefault}{\mddefault}{\updefault}{\color[rgb]{0,0,0}0}%
}}}
\put(4801,-1561){\makebox(0,0)[b]{\smash{\SetFigFont{8}{7.2}{\rmdefault}{\mddefault}{\updefault}{\color[rgb]{0,0,0}Decoding regions}%
}}}
\put(4951,-286){\makebox(0,0)[lb]{\smash{\SetFigFont{8}{7.2}{\rmdefault}{\mddefault}{\updefault}{\color[rgb]{0,0,0}Correct answer: $b_t = -1$}%
}}}
\put(3901,-1186){\makebox(0,0)[b]{\smash{\SetFigFont{8}{7.2}{\rmdefault}{\mddefault}{\updefault}{\color[rgb]{0,0,0}-2}%
}}}
\put(4351,-1186){\makebox(0,0)[b]{\smash{\SetFigFont{8}{7.2}{\rmdefault}{\mddefault}{\updefault}{\color[rgb]{0,0,0}-1}%
}}}
\put(5251,-1186){\makebox(0,0)[b]{\smash{\SetFigFont{8}{7.2}{\rmdefault}{\mddefault}{\updefault}{\color[rgb]{0,0,0}1}%
}}}
\put(5701,-1186){\makebox(0,0)[b]{\smash{\SetFigFont{8}{7.2}{\rmdefault}{\mddefault}{\updefault}{\color[rgb]{0,0,0}2}%
}}}
\put(6001,-961){\makebox(0,0)[lb]{\smash{\SetFigFont{8}{7.2}{\rmdefault}{\mddefault}{\updefault}{\color[rgb]{0,0,0}$|{\cal B}|=5$}%
}}}
\put(6001,-1186){\makebox(0,0)[lb]{\smash{\SetFigFont{8}{7.2}{\rmdefault}{\mddefault}{\updefault}{\color[rgb]{0,0,0}five possible}%
}}}
\put(6001,-1411){\makebox(0,0)[lb]{\smash{\SetFigFont{8}{7.2}{\rmdefault}{\mddefault}{\updefault}{\color[rgb]{0,0,0}output letters}%
}}}
\put(5851,-2686){\makebox(0,0)[lb]{\smash{\SetFigFont{8}{7.2}{\rmdefault}{\mddefault}{\updefault}{\color[rgb]{0,0,0}Adjusted
	by $-\lambda F(b_{t-1})$}%
}}}
\put(4351,-2461){\makebox(0,0)[rb]{\smash{\SetFigFont{8}{7.2}{\rmdefault}{\mddefault}{\updefault}{\color[rgb]{0,0,0}Actual control $U'_t$}%
}}}
\put(4426,314){\makebox(0,0)[b]{\smash{\SetFigFont{8}{7.2}{\rmdefault}{\mddefault}{\updefault}{\color[rgb]{0,0,0}$X_{t+1} - \lambda X_{t} = U'_{t} + W_{t}$}%
}}}
\put(5851,-2911){\makebox(0,0)[lb]{\smash{\SetFigFont{8}{7.2}{\rmdefault}{\mddefault}{\updefault}{\color[rgb]{0,0,0}at controller to compensate}%
}}}
\end{picture}
\end{center}
\caption{How to communicate the channel outputs through the plant with
state observations only. The controller restricts its main control
signal to be either zero, or calculated with a delay of $1$ time unit.
It is adjusted by $-\lambda F(b_{t-1})$ to eliminate
the effect of the past communication. The final control signal applied
is shifted slightly to encode which $b_t$ was received. The observer
uses the past $b_0^{t-1}$ to align its decoding regions and then reads
off $b_t$ by using $X_{t+1} - \lambda X_t$.}
\label{fig:latticedecoder}
\end{figure}

The approach is illustrated in Figure~\ref{fig:latticedecoder} and
detailed in \cite{ControlPartI, SahaiCDC04}. The goal is to
communicate the noisy channel outputs back to the observer so it can
guide the controller. The main control signal $U_t$ is zero except
when directed by the observer as depicted in
Section~\ref{sec:supervision}. The actual applied control is 
\begin{equation} \label{eqn:dancingcontrol}
U'_t = U_t + F(b_t) - \lambda F(b_{t-1})
\end{equation}
where $F(b_t) = 3 \Gamma_u b_t$ and $\Gamma_u = \Omega + (\lambda +
1)\Gamma$ where $\Gamma$ is the bound on the noise in state
observation. If $X_o(t)$ is the boundedly noisy observation of the
state $X_t$, the observer can compute $X_o(t) - \lambda X_o(t-1)$
which is going to be $U'_{t-1} + W_{t-1}$ corrupted by a bounded noise
of at most $\Gamma$ from $X_o(t)$ and $\lambda \Gamma$ from $\lambda
X_o(t-1)$. The total noise in observing $U'_{t-1}$ is thus $\Gamma_u$
and so with this choice of $F$, the channel output $b_t$ can be
unambiguously determined by the observer.

If we assume that there is a finite channel output alphabet, then
because of the cancellation of the past $F(b_{t-1})$, this dance has
no lasting impact on the state and thus can be ignored for stability
considerations.

\subsection{Nonuniform sampling and control choice} \label{sec:nonuniform}

As in \cite{ACC00Paper}, the controller must keep track of a total
window of uncertainty in which the state lies. The observer also
tracks this window so as to communicate to the controller. Call the
current uncertainty window $\Delta_t$. At a sampling time, the
observer divides this window into $\exp(nR)$ (large) different regions
and sees where the state observation is. Each region is assigned an
infinitely long iid codeword drawn uniformly from the channel input
alphabet according to the optimizing $\vec{q}$. Letters from this
codeword are transmitted on the noisy channel until the state is
resampled.

If the controller faced an uncertain window of $\Delta_t$ at time $t$,
then without any information, the window will grow according to:
\begin{equation} \label{eqn:deltaevol}
\Delta_{t+1} = \lambda \Delta_t + \Omega
\end{equation}
If the controller learns correctly which region the observation lies
in, then the retrospective uncertainty shrinks to at most $\Gamma +
\Delta_t \exp(-nR)$. If this occurs $T$ time steps in the future, the 
resulting uncertainty is at most
\begin{equation} \label{eqn:bigdeltaevol}
\Delta_{t+T} = \lambda^{T}(\exp(-nR)\Delta_t + \Gamma + \frac{\Omega
  \lambda}{\lambda - 1})
\end{equation}
and knowledge of the window allows the controller to apply a ``true''
control that will center this uncertainty window around $0$.

As in \cite{ACC00Paper}, it is how the randomness in $\lambda^T$
compares to $\exp(nR)$ that will determine whether or not the system
is $\eta$-stable in closed loop.

\subsection{Supervision through the noiseless channel outputs} \label{sec:supervision}

Due to the ``dancing'' of Section~\ref{sec:dancing}, the observer
knows what channel outputs have been received so far. Thus, it can
monitor whether the channel outputs are good enough for the controller
to successfully decode which region the state was in at the last state
sampling time. When that occurs, it sends a $1$ over the noiseless
channel and otherwise it sends a zero. Thus, the random variable $T$
is the time till successful decoding of an infinitely long codeword
chosen out of $\exp(nR)$ possible codewords. So far, $nR$ was just an
arbitrary constant. To interpret it, think of $\ln \lambda < R <
\frac{E_0(\eta + \epsilon)}{\eta + \epsilon}$. Choose $\rho$ so that
$R = \frac{E_0(\rho)}{\rho}$. Think of $n$ as a constant that is a
large multiple of $k$.

At this point, the detailed analysis in \cite{OurUpperBoundPaper}
explains what happens. The key points are summarized here.
\begin{itemize}
 \item[a.] The $T$ represent the inter-sampling times of the
       state. This process is iid across time since the codewords are
       randomly chosen and the noisy channel is memoryless.
 \item[b.] If $\rho \leq 1$ works, then the scheme described above
       works without modification and the random variable $T$ is
       strictly dominated by $n + \widetilde{T}$ where the $\widetilde{T}$ is
       a geometric random variable with probability of failure given
       by $\exp(-E_0(\rho))$.
 \item[c.] If $\rho > 1$ is required, then the scheme above must be
       modified to use list decoding among the top $l = \lceil \rho
       \rceil$ entries with the next $\lceil \log_2 l \rceil$
       noiseless channel uses used to disambiguate the list. This is a
       small overhead that is negligible when $nR$ is large. With
       that modification, everything from (b) still holds in that 
       $T \leq n + \lceil \log_2 l \rceil k + \widetilde{T}$.
\end{itemize}

At this point, we give the key ideas in the proof: 

Using (b), we can rewrite (\ref{eqn:bigdeltaevol}) to get
\begin{eqnarray*}
\Delta_{t+T} 
& = & 
\lambda^{T}(\exp(-nR)\Delta_t + \Gamma + \frac{\Omega \lambda}{\lambda - 1}) \\
& \leq & 
\lambda^{n + \widetilde{T}}(\exp(-nR)\Delta_t + \Gamma + \frac{\Omega \lambda}{\lambda - 1}) \\
& = & 
\lambda^{\widetilde{T}}(\lambda^{n}\exp(-nR)\Delta_t + \Gamma \lambda^n + 
  \frac{\Omega \lambda^{1 + n}}{\lambda - 1}) \\
& < & 
\lambda^{\widetilde{T}}(\exp\left[-n (R - \ln \lambda)\right]\Delta_t +
\Gamma \lambda^n + 
  \frac{\Omega \lambda^{1 + n}}{\lambda - 1})
\end{eqnarray*}

Call this upper bound $\bar{\Delta}_{t+T}$ so:
\begin{equation} \label{eqn:boundevolution}
\bar{\Delta}_{t+T} = \lambda^{T-n}(\exp\left[-n (R - \ln
  \lambda)\right]\bar{\Delta}_t + \Gamma \lambda^n + 
  \frac{\Omega \lambda^{1 + n}}{\lambda - 1})
\end{equation}
where we assume that the next successful control application after
time $t$ occurs at time $t+T$. 

It is worthwhile interpreting the random variable $\widetilde{T}$ as
the inter-arrival time for a packet-erasure channel and $\exp\left[n
  (R - \ln \lambda)\right]$ as the number of possible messages carried
by a single packet. Since $R > \ln \lambda$, the size of the packet
can be made as large as we want by picking a large\footnote{Of course,
  picking a large $n$ also makes the effect of the disturbance and
  observation noise greater, but it remains bounded no matter how
  large $n$ is and that is all that matters when it comes to the
  existence of $\eta$-moments.} $n$. Viewed this way, the uncertainty
window $\Delta$ is very likely to shrink from one arrival to the next
if it starts out large.

Asymptotically, the problem behaves like the very large packet
erasure-channel case with a packet-erasure probability of
$\exp(-E_0(\rho))$. For $n$ large enough, the closed-loop system will
have an $\eta$-moment as long as $\lambda^\eta \exp(-E_0(\rho)) < 1$
or by taking logs: $E_0(\rho) > \eta \ln \lambda$ which is true by
assumption.

\subsection{Event-based sampling is simpler than ``dancing''}
\label{sec:nodancing}

While the above three-part strategy makes for an easy proof, it
invites criticism for being very unrealistic:
\begin{itemize}
 \item The ``dancing'' of the plant seems to be excessive. When the
   channel output alphabet is large, this gives a tremendous
   performance hit as the channel outputs are communicated back
   through the plant. This also does not work if there is a continuous
   or other infinite channel output alphabet.

 \item The ``observer'' here essentially has to run a copy of the
   channel decoder in the controller as well as interpret the plant
   dancing. There is thus duplication of effort in the whole
   system. 
 \end{itemize}

To overcome this, we just have to notice that essentially all the
controller has to be told is when to stop decoding this block and
when it is time to move on to the next block. We can view successful
decoding as an ``event'' that defines when the observer will next
``sample'' the state process for the purpose of generating channel
inputs. The above strategy uses the noiseless feedback of channel
outputs through the plant to decide when successful decoding as
occurred --- however there is a simpler way to see this. {\em Whenever
the controller is successful in decoding the channel outputs, it will
be able to apply a control that drives the state to the neighborhood
of zero!} 

\begin{figure}
\begin{center}
\includegraphics[width=5in]{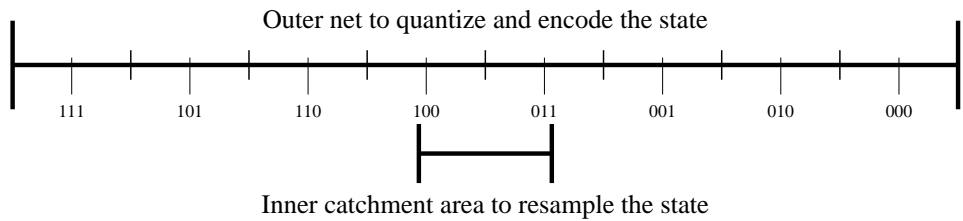}
\end{center}
\caption{The observer maintains two ``nets'' that it uses to catch the
  state. The outer ``quantization net'' represents everywhere the
  controller believes the state could possibly be based on the
  information it is known to have received so far. At sampling time,
  it is partitioned into $e^{nR}$ uniformly sized boxes and the box
  corresponding to the true state is used to pick a message to send
  over the noisy channel. The inner net (of size $\bar{\Delta}_{t+T}$
  at time $T$ when the last sampling event occurred at time $t$)
  represents where the controller could drive the state given that
  it was able to successfully decode the current message block. Both
  nets grow steadily exponentially at rate $\approx \lambda$ per unit
  time, and shrink suddenly whenever the controller is able to
  successfully decode and have this success confirmed through a
  fortified bit.} 
\label{fig:innerouter}
\end{figure}

This gives rise to modifications to the control strategy that do not
rely on the finite channel output alphabet assumption:
\begin{itemize}
 \item The controller no longer encodes the channel outputs into the
       control signals by making the plant dance. 

 \item Instead of waiting for a ``go ahead'' signal from the
       observer through the fortification channel, the controller
       applies the control based on its tentative best decision so
       far. If it gets negative feedback through the fortification
       channel, it backs-out the tentative control\footnote{By adding
         $-\lambda U_{t-1}$ to its new control when it wants to
         countermand its previous control.} it had applied and crosses
       the corresponding message off the list of possibilities. It
       then moves on to the new top-scoring message and applies the
       corresponding control until it receives positive feedback.  
 
 \item Instead of using the fortification channel to tell the controller
       when to stop decoding and apply a control, the observer uses
       the fortification channel to simply tell the decoder when it
       has gotten the right answer. This can be observed at the
       observer by seeing the plant state move into the
       vicinity\footnote{By vicinity, we simply mean that the
       observed state lands within a box of size $\bar{\Delta}_{t+T}$ of the 
       origin when considering time $t+T$.} of the origin. 

 \item Otherwise, the observer remains essentially as before when
       it comes to sampling the controlled state and determining what
       infinitely-long random codeword to start sending
       incrementally. 
\end{itemize}

In terms of analysis, there are only two remaining concerns:
\paragraph{False control performance hit}
Before a true control action is confirmed, the controller might try
some different false controls. However, a false control can at worst
move the state to twice the outer limits of the current uncertainty
window. This happens when the true control should be dealing with one
extreme endpoint and the false applied control corresponds to the
other endpoint. However, this factor of $2$ just changes the bound on
the $\eta$-th moment by a factor of $2^\eta$. It will not make a
moment infinite if it were finite before.

\paragraph{List decoding}
In the original scheme above that was made to leverage the purely
channel coding derivation in \cite{OurUpperBoundPaper}, list-decoding
was required to disambiguate some remaining uncertainty. This caused
this small constant $(1 + \log_2 l)k$ term to show up along with the
original $n$. But since $l$ is a constant that does not depend on $n$,
the $\log_2 l$ is not necessary for the result to hold. Everything
will work just as well asymptotically even if we used a ``unary
encoding'' and used $(l+1)k$ time-slots for the
list-disambiguation. This is essentially what the new observer does
since it just keeps ruling out the seemingly most likely codewords
until the true one is applied. 

Consequently, the event-based sampling approach that uses the
fortification to give simple positive and negative supervision to the
controller will achieve the focusing-bound performance. \hfill $\Box$

\section{Conclusion} \label{sec:conclusion}
\begin{figure}[tbph]
\begin{center}
\begin{minipage}[t]{3.3in}
\includegraphics[width=3.2in]{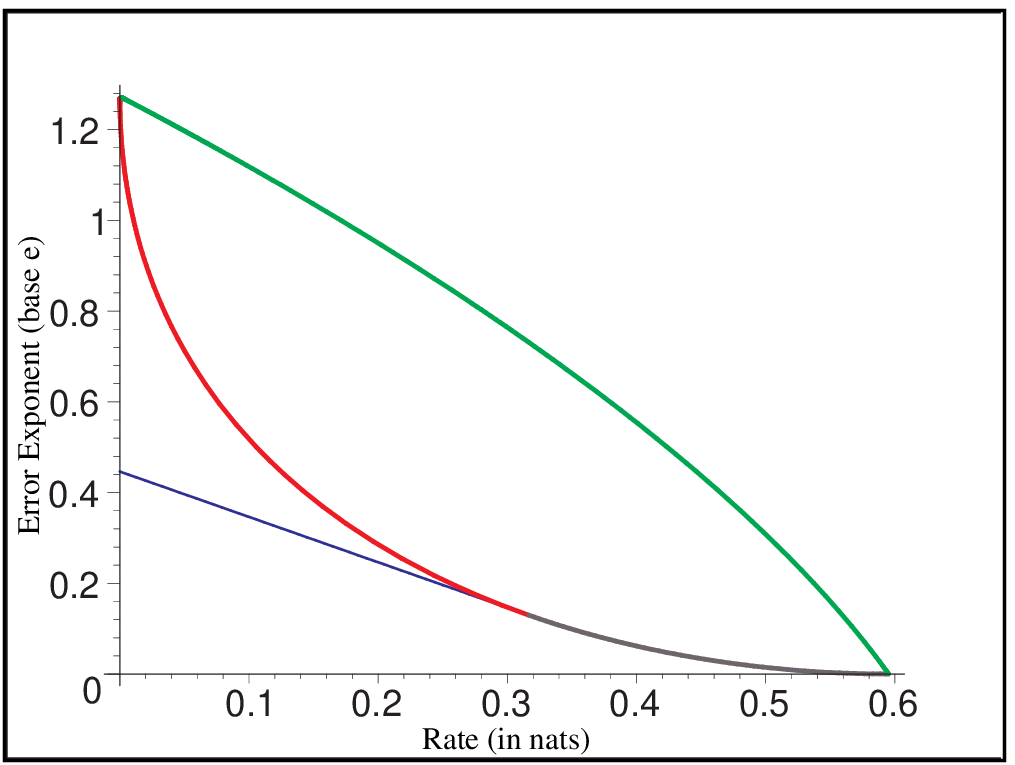}
\end{minipage}
\hfill
\begin{minipage}[t]{3.3in}
\includegraphics[width=3.2in]{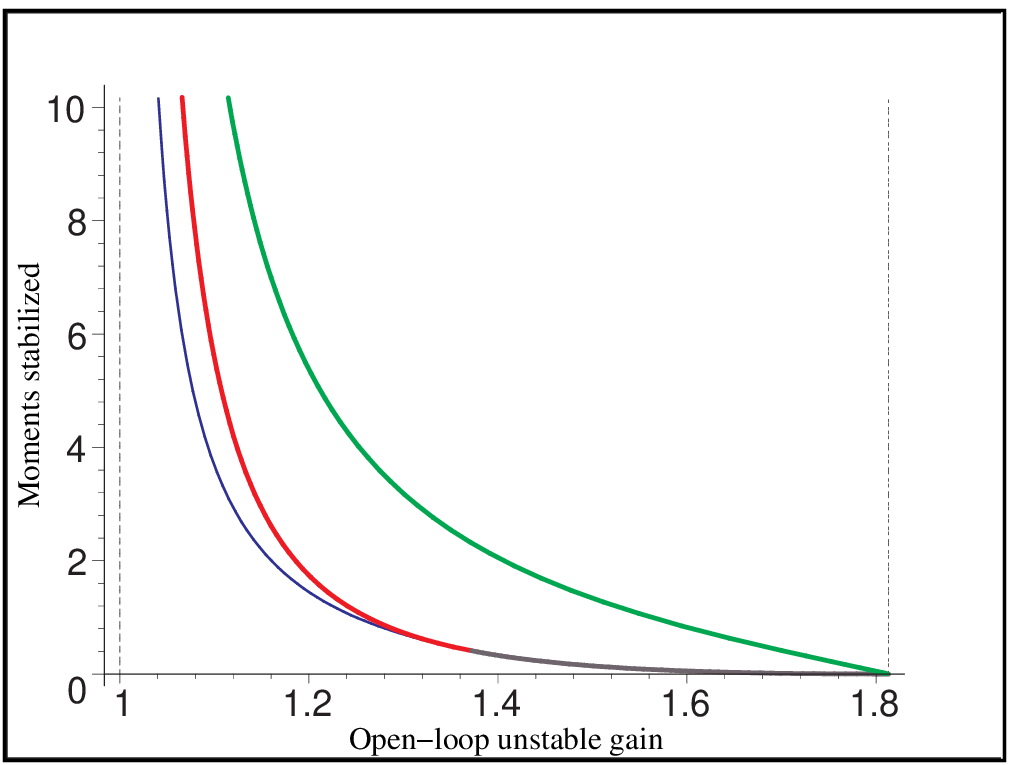}
\end{minipage}
\end{center}
\caption{The sphere-packing and focusing bounds for the binary
  symmetric channel with crossover probability $0.02$. On the left,
  reliability is plotted relative to rate, while on the right, the the
  stabilized moment is plotted relative to the unstable gain
  $\lambda$. The focusing bound is the highest and bounds what is
  possible with feedback, the sphere-packing bound is below and bounds
  the reliability for communication without feedback. The random
  coding error exponent matches that bound at high-rates, but falls
  below at low rates. Since stabilization is equivalent to
  communication with feedback, the upper bound is the target.}
\label{fig:twobounds}
\end{figure}

The main result of this paper is that a little bit of noiseless
feedback used for supervision can be used to leverage the
effectiveness of the noisy feedback. The difference in what can be
achieved is illustrated in Figure~\ref{fig:twobounds}. With even a
tiny bit of noiseless feedback, we can achieve performance
corresponding to the upper curves.

The theorem as stated is asymptotic in nature and it might appear
troubling to have to use large $n$ and thereby significantly sacrifice
performance just in order to stabilize larger moments. However, by
applying the ``supervisory'' idea again, no such sacrifice is needed.
The control strategy of \cite{Allerton05Control} can be applied and
most of the time, it will hold the plant very close to the origin.
Rarely, the state will wander out of a particular target ball. At that
point, the observer can use the noiseless channel uses to declare an
``emergency'' and switch to the control strategy given in this paper.
Once the state has returned to a moderate sized neighborhood of the
origin, the state of emergency can be lifted and the control strategy
can return to that of \cite{Allerton05Control}. By making the target
ball large, emergencies will only occur rarely and thus will not
impact average performance. However, the emergency mode will ensure
that as many higher moments of the state will exist as is possible for
the particular channel in question.

\bibliographystyle{IEEEtran}
\bibliography{IEEEabrv,./MyMainBibliography} 
\end{document}